**Title: Establishing phase diagram for the band engineering in *p*-type PbTe/SnTe from elementary electronic structure understanding**

*Xiaojian Tan, Guoqiang Liu,* Jingtao Xu, Hezhu Shao, Haochuan Jiang, and Jun Jiang**

Dr. X. Tan, Prof. G. Liu, Dr. J. Xu, Dr. H. Shao, Prof. H. Jiang, Prof. J. Jiang
Ningbo Institute of Materials Technology and Engineering, Chinese Academy of Sciences, Ningbo 315201, China
E-mail: liugq@nimte.ac.cn, jjun@nimte.ac.cn

Band engineering is an important mechanism to increase the thermopower of thermoelectric materials by reconstructing the band structure near Fermi level. PbTe and SnTe are the most representative systems in which band engineering were achieved by various dopants. Starting with the elementary understanding of the band structures, we established the phase diagram for the band engineering in *p*-type PbTe/SnTe by constructing an *s*-*p* bonding model. We show that the effects of band tuning are mainly determined by an inherent parameter of doping element: the *s* orbital energy level. With the phase diagram, all the related experimental observations can be consistently explained, moreover, undiscovered effective dopants become foreseeable. Our study discovers an applicable criteria to pick up proper dopants from the periodic table directly, and the analytical method can be adopted to more thermoelectric materials.

## 1. Introduction

Waste heat recovery is an environmentally-friendly way to preserve resources and increase energy efficiency. Thermoelectric materials can directly convert heat into electricity by using temperature difference to induce current flow.[1] Thermoelectric technology in commercial applications, however, has been limited because of the low conversion efficiency. The thermoelectric figure of merit $ZT$ is described by $ZT = S^2\sigma T/(\kappa_e+\kappa_{ph})$, where S, σ, T, $\kappa_e$, $\kappa_{ph}$ are the Seebeck coefficient, electrical conductivity, absolute temperature, electronic and lattice thermal conductivity, respectively.[2] The key problem here is that these properties are strongly coupled, and thus they could not be individually optimized. When the electron-phonon coupling is weak, as in most of thermoelectric materials, *ZT* can be improved by reducing the lattice thermal conductivity $\kappa_{ph}$ or increasing the power factor $S^2\sigma$ quasi-independently. The pure electron property $S^2\sigma$ is dominated by the details of band structure and scattering mechanism. Band engineering is a mechanism to increase $S^2\sigma$ by reconstructing the geometry of band structure near Fermi level. As for a nonmagnetic material, band engineering includes two main types: band convergence (BC)[3] and resonant level (RL),[4] as summarized in **Figure 1**.

The thermoelectric quality $\beta$, which indicates the maximum attainable *ZT* for a given system, is in proportion to the number of valley degeneracy, $\beta \propto N_v$.[5,6,7] Increasing $N_v$ can significantly improve S while the electrical conductivity s is relatively less influenced. As for the two valley structure shown in Fig. 1(a), if the distance between light and heavy valence bands is reduced to several meV by chemical doping or temperature increasing, the power factor will be enhanced. RL is a distortion of density of state (DOS). As shown in Fig. 1(b), a DOS 'hump', which could be introduced by chemical doping, appears around Fermi level. As discussed by Mahan *et al*,[8] such enhanced DOS peak can improve *ZT*, depending on its sharpness and position.

PbTe is the most successful example of band engineering.[9,10] BC in PbTe was achieved by various dopants, such as Mg,[11] Ca/Ba,[12] Sr,[13]Mn,[14] Cd,[15]and Eu.[16] On the other hand, RL in PbTe is much more element-selective. RL was reported only in Tl-doped *p*-type PbTe,[4] and a few transition metals doped *n*-type PbTe.[17,18,19] Band engineering, combined with the optimization of lattice thermal conductivity, improved the *ZT* of PbTe from 0.6 to 2.2–2.5.[3,13,20] SnTe has similar band structure to PbTe, and it exhibits similar band engineering effects to PbTe.[21,22,23,24,25] BC in SnTe was achieved by Mg,[26] Ca,[27] Mn,[28,29,30,31] Cd,[32,33] or Hg[34] doping, and RL by In doping.[35,36] In Mn-Cu codoped SnTe, the highest *ZT* of 1.6 for SnTe based materials was reported.[37]

In general, the studies of band engineering were led by experimental researches. Despite the experimental successes, the microscopic mechanisms behind band tuning were not well understood. For example, the elements Mg, Mn, Cd, and Eu have similar doping effects in PbTe.[11,14,15,16] Since these elements are from different groups and different periods of the periodic table, it is difficult to tell what kind of characteristic they share. Experimentally, numerous efforts had been paid to find out these effective dopants.[5,38,39,40,41,42,43,44,45,46,47,48] If the mechanism of band engineering can be disclosed, we might find a directly way to pick up effective dopants, and the burdensome works could be relieved.

In this work, we start from the analysis of the band structures for PbTe/SnTe. It is discovered that the valence band dispersion is mainly from the anti-bonding of cation-*s* and anion-*p* orbitals. By constructing an *s*-*p* bonding model, we show that the effects of band tuning are essentially determined by an inherent parameter of doping element: the *s* orbital energy level, $\varepsilon_s$. Taking $\varepsilon_s$ as the variable, the phase diagram of band engineering in *p*-type PbTe/SnTe is established. This phase diagram not only explains the related experimental observations, but also predicts some undiscovered effective dopants. For example, Be is predicted to be a very efficient dopant for BC in SnTe, and the prediction is further confirmed by the band structure calculations. Our work opens a new thought to design band engineering for thermoelectric materials, not only for PbTe/SnTe but also for more systems.

## 2. Elementary Band Structures of PbTe/SnTe

### 2.1. Valence Band and *s*-*p* Anti-Bonding State

To understand the BC and RL mechanisms in PbTe/SnTe, we start from their band structures. The band structures of PbTe/SnTe were firstly calculated in the 1960s.[49,50] It was realized that these tin and lead chalcogenides possess many abnormal electronic properties. The gaps of lead chalcogenides increase with temperature and decrease with pressure, in contrast to most II-VI and III-V semiconductors.[51,52,53] The gaps of lead chalcogenides exhibit a nonmonotonic order: those are 0.19, 0.165, and 0.285 eV for PbTe, PbSe and PbS, respectively.[54] The valence-band maximum (VBM) and conduction-band minimum (CBM) of lead chalcogenides appear at L point. The VBM and CBM of PbTe were denoted to be $L_6^+$ and $L_6^-$ states, respectively, while they were reversed in density functional theory (DFT) calculations.[55] This deficiency was overcome by the more sophisticated methods, such as hybrid functionals and GW approach.[56,57,58,59,60,61] In this work, we focus on the formations of the $L$-$\Sigma$ two-valley structure and the band gap, which are directly related to the band engineering in *p*-type PbTe/SnTe.

**Figure 2** displays the band structures and partial DOS for PbTe. Although the spin-orbit (SO) coupling effect was reported to significant in tin and lead chalcogenides,[49,50] its strength is still much weaker than chemical bonding and it only influences some high symmetry ***k***-points. This work would like to present a picture as simple as possible, and therefore the SO coupling is ignored in the calculations of Fig. 2, from which we shall extract a simple valence band model. In the rest band structure calculations, the SO coupling is included to examine the predictions of our theory.

Fig. 2(a) presents the partial DOS for PbTe. As may be seen, the majorities of Te-*s*, Pb-*s*, Te-*p*, and Pb-*p* orbitals are located around –11, –8, –2 and 2 eV, respectively. This location order is determined by their atomic energy levels. From –4 to 0 eV, DOS is dominated by Te-*p* orbital, while Pb-*s* and Pb-*p* states exhibit two peaks around 0 eV and –3 eV, respectively. It is easy to know that the two peaks signify the anti-bonding state of Te-*p* and Pb-*s*, and the bonding state of Te-*p* and Pb-*p*.

Fig. 2 (b-c) present the character bands for Pb-*s* and Te-*p* orbitals. Interestingly, all the Pb-*s* character nearly fall on the first valence band. Fig. 2(b) shows that the first valence band contains more Te-*p* state. We then may conclude that the behavior of first valence band is mainly determined by the anti-bonding state of Pb-*s* and Te-*p* orbitals. In fact, this finding is not surprising. The anti-bonding (bonding) state is always pushed up (down) by the interactions between two orbitals. In a bundle of bands, the anti-bonding or bonding state between two nearest neighbors is always distributed at the edges. Thus a simplest model to describe the valence band of PbTe needs to include the Pb-*s* and Te-*p* orbitals.

### 2.2. Tight-Binding Model for Valence Band Dispersion

According to the rock-salt structure of PbTe, the ***k***-space tightbinding (TB) Hamiltonian can be given as:

$$H(\vec{k}) = \begin{pmatrix} \varepsilon_{Pb-s} & h_{sx} & h_{sy} & h_{sz} \\ h_{sx} & \varepsilon_{Te-p} & 0 & 0 \\ h_{sy} & 0 & \varepsilon_{Te-p} & 0 \\ h_{sz} & 0 & 0 & \varepsilon_{Te-p} \end{pmatrix}.$$

For the convenience of discussion, the Hamiltonian can be equivalently written as:

$$H(\vec{k}) = \begin{pmatrix} \Delta\varepsilon & h_{sx} & h_{sy} & h_{sz} \\ h_{sx} & 0 & 0 & 0 \\ h_{sy} & 0 & 0 & 0 \\ h_{sz} & 0 & 0 & 0 \end{pmatrix}. \quad (1)$$

The matrix elements are

$$\Delta\varepsilon = \varepsilon_{Pb-s} - \varepsilon_{Te-p}$$

$$h_{sx} = 2t_{sp\sigma}\sin(k_x/2)$$

$$h_{sy} = 2t_{sp\sigma}\sin(k_y/2)$$

$$h_{sz} = 2t_{sp\sigma}\sin(k_z/2) \quad (2)$$

where $\varepsilon_{Pb\text{-}s}$ and $\varepsilon_{Te\text{-}p}$ are the on-site energies for Pb-*s* and Te-*p* orbitals, $t_{sp\sigma}$ is the hopping integral for the *spσ* bonding. The orbital interaction is considered up to the first nearest neighbors. Thus the model includes two undecided parameters, Δε and $t_{sp\sigma}$.

In principle, the on-site energies of certain orbitals are defined as their atomic energy levels, which can be looked up in textbooks. **Table 1** presents the atomic energy levels for the orbitals of tin and lead chalcogenides from Harrison's book.[62] In a real system, the on-site energies are affected by the crystal field, and thus they will deviate from the atomic energy levels. In spite of the deviation, the on-site energies can be approximately estimated from the listed energy levels. Table 1 shows an order of $\varepsilon_{\text{Te-}s} < \varepsilon_{\text{Pb-}s} < \varepsilon_{\text{Te-}p} < \varepsilon_{\text{Pb-}p}$, which is consistent with the calculated DOS shown in Fig. 2(a). The hopping integral $t_{sp\sigma}$ is difficult to estimate, for it is sensitive to the bond length and bond angle. To determine the two parameters, we need to fit the DFT band structure.

The TB Hamiltonian is a $4 \times 4$ matrix, which gives 4 eigenvalues at each $\boldsymbol{k}$-point. As shown in Fig. 2, the valence band is characterized by an anti-bonding state, informally written as: $c_1\varphi_s - c_2\varphi_p$ with $|c_2| > |c_1| \gg 0$. In principle, the valence band can be picked up by reading the eigenvectors. In fact, the antibonding state gives the largest eigenvalue at each $\boldsymbol{k}$-point, and it is not necessary to check the eigenvectors.

**Figure 3**(a) displays the fitted valence band of PbTe, compared with the DFT band structure. Although the deviation between two methods is obvious, the TB valence band reproduces the two-valley structure. The deviation is mainly from the ignoring of the *p*-*p* interactions between the first nearest neighbors in the TB model. As shown in Fig. 2(c-d), the Pb-*p* character appears on some $\boldsymbol{k}$-points of the first valence band, although it mainly falls on the third valence band. The hopping term for *p*-*p* interactions can be written as: $2t_{pp\sigma}\cos(\vec{k}/2)$. It is easy to find that Pb-*p* character does not appear on $L$ (π, π, π) point, but has a strong influence on the $\Gamma$ (0, 0, 0) point. In the system where the *p*-*p* interactions are absent, such as in MgTe, the fitting can be significantly improved. In next section, we will show that the TB band for MgTe is in good agreement with the DFT result. Our purpose is to present a model, as simple as possible, to understand the general behaviors of lead and tin chalcogenides. If the *p*-*p* interactions are included in the TB model, the essential feature of valence band might be hidden in the increased information.

Fig. 3(b) displays the calculated energy difference of $\Delta E_{L\text{-}\Sigma}$ as a function of $\Delta\varepsilon$ and $t_{sp\sigma}$. As for all the lead and tin chalcogenides, $\Delta\varepsilon$ is negative. In the next section, we will discuss the situation of $\Delta\varepsilon > 0$. At $\Delta\varepsilon < 0$, the figure shows that $E_L$ is always higher than $E_\Sigma$ for different $t_{sp\sigma}$, *i.e.*, the two-valley structure is mainly driven by $\Delta\varepsilon$. This is not surprising because $\Delta\varepsilon$ is the diagonal element of the Hamiltonian.

The TB model can simply explain some abnormal behaviors of lead and tin chalcogenides.

1) The VBMs appear at L point in the Brillouin zone. At $L$ (π, π, π) point, the hopping terms $2t_{sp\sigma}\sin(\vec{k}/2)$. take their maxima, *i.e.*, the largest effective *s*-*p* interaction occurs at $L$ point. Thus the highest eigenvalue of anti-bonding state is at $L$ point. In the work of Wei *et al.*,[63] this *s*-*p* interaction was called *s*-*p* level repulsion. The authors stated that the *s*-*p* repulsion is strong at $L$ point, and absent at $\Gamma$ point. The expression of our model explains their statements straightforwardly.

2) $\Delta E_{L\text{-}\Sigma}$ increases along PbTe-PbSe-PbS.[55] As shown in Table 1, the distance between anion-*p* and Pb-*s* orbitals is decreasing along the serials of PbTe-PbSe-PbS. The effective *s*-*p* interaction will increase as $|\Delta\varepsilon|$ decreases, leading to the amplifying ***k***-space dispersion. Therefore the distance between $E_L$ and $E_\Sigma$ is increasing along PbTe-PbSe-PbS, as illustrated in Fig. 3(b).

3) The band gaps of lead chalcogenides take an abnormal order of $E_{g\text{-PbSe}} < E_{g\text{-PbTe}} < E_{g\text{-PbS}}$.[54] The measured band gaps for lead chalcogenides is listed in Table 1. As shown in Fig. 2(b), the band gap of lead chalcogenide is roughly scaled by the positions of Pb-*p* and anion-*p* orbitals. With the increasing distance between Pb-*p* and anion-*p* orbitals along PbTe-PbSe-PbS, the band gap should increase. On the other hand, the increasing valence band dispersion along PbTe-PbSe-PbS widens the bandwidth, and then decreases the band gap. The competition of the two effects causes the nonmonotonic order.

**3. Phase Diagram of Band Engineering in *p*-Type PbTe/SnTe**

We have shown that the valence band of PbTe/SnTe is dominated by the *s*-*p* anti-bonding state, whose strength mainly depends on the atomic energy levels of $\varepsilon_s$ and $\varepsilon_p$. When a dopant is introduced to PbTe/SnTe, it is conceivable that the doping effect on the valence band are also related to the two energy levels. **Figure 4** is the phase diagram for the band engineering in *p*-type PbTe/SnTe established from our model. We will show that the band engineering effect is a function of $\varepsilon_{D\text{-}s}$, the *s* orbital energy of a given dopant.

### 3.1. $\varepsilon_{D\text{-}s} > \varepsilon_{Te\text{-}p}$: Band Convergence

Researchers have achieved a series of successful BC in *p*-type PbTe and SnTe.[11,12,13,14,15,26,27,28,29,30,31,32,33,34] For example, Pei *et al.* reported BC effects in the alloy of PbTe and MgTe, where the average *ZT* is improved to 0.9 with a peak *ZT* of 1.7.[11] Biswas *et al.* synthesized nano-structured PbTe-CaTe/BaTe alloy, and they observed increased power factor by BC effects besides the decreased lattice thermal conductivity.[12] By alloying MnTe and CdTe, Pei *et al.* increased the peak *ZT* of PbTe to 1.6 and 1.7, respectively.[14,15] Recently, Tan *et al.* reported a significantly improved *ZT* of 2.5 at 923 K in PbTe-SrTe alloy,[13] while the enhanced thermoelectric performance was attributed to the BC effects and reduced lattice thermal conductivity.

In SnTe, Banik *et al.* achieved BC optimization by alloying MgTe, and a peak *ZT* of 1.2 at 860 K is achieved.[26] Similar BC effects were reported in Ca-doped SnTe, and the maximum *ZT* is up to 1.35 at 870 K.[27] Tan *et al.* reported that BC effects in Hg-doped SnTe lead to a high *ZT* of 1.35 at 910 K.[34] BC effects in MnTe alloying SnTe were simultaneously observed by several independent groups,[28,29,30,31] and the maximum *ZT* is enhanced to 1.3–1.4 around 900 K. In Cd-doped SnTe, BC effects increases maximum *ZT* to about 1.0 around 800 K.[32,33] In Mn-Cu codoped SnTe, the highest *ZT* of 1.6 for SnTe based materials was reported.[37]

**Figure 5** summarized the improved Seebeck coefficient for the above reported BC doping in PbTe/SnTe. It may be seen that BC doping can significantly enhance the seebeck coefficient relative to the Pisarenko curves. Since SnTe possesses larger energy separation (0.35 eV) between light and heavy valence bands than PbTe (0.17 eV),[21,22,23] higher doping concentration in SnTe is required to reach similar BC effects in PbTe.

In **Table 2**, we list some elements with their s orbital energy levels. In the listed elements, Cd, Mg, Mn, and Ca were reported to be able to induce BC in PbTe and SnTe.[11,12,14,15,26,27,28,29,30,31,32,33] One may find that the s energy levels of these elements are all higher than that of Te-*p* orbital. Taking Mg doping as an example, we explain the mechanism of BC. Firstly, we consider a MgTe compound in the rock-salt structure.

Fig. 3(c) presents the band structures calculated by DFT and our TB model. Since the Mg-*s* orbital is higher than the Te-*p* orbital, the Mg-*s* character in the Te-*p* dominated region indicates an *s*-*p* bonding state, which can be informally written as: $c_s\varphi_s + c_p\varphi_p$ with $|c_p| > |c_s| \gg 0$. Fig. 3(c) shows that the *s*-*p* bonding state by TB fitting is in very good agreement with that of DFT. In contrast to PbTe, the *p*-*p* interaction is absent in MgTe. That is the reason why the agreement between the *s*-*p* model and DFT is much better in MgTe than in PbTe. It is noticeable that the *s*-*p* bonding state is the reversal of anti-bonding state.

In Fig. 3(b), the energy difference $\Delta E_{L\text{-}\Sigma}$ at $\Delta\varepsilon > 0$ is for the *s*-*p* bonding state. One may see that the $L$ point is always lower than the $\Sigma$ point in a bonding state. When a defect *s*-*p* bonding state is introduced to PbTe/SnTe by Mg doping, $E_L$ and $E_\Sigma$ will be downward shifted, for the anti-bonding state of the bulk material is partly offset by the dopant. Owing to the decrease of band dispersion, the band gap is enlarged and the distance between the heavy and light bands is reduced.

The above picture is a phenomenological explanation. Here we give a more analytical discussion. For the readers who are not familiar with TB theory and vector space, this discussion can be skipped, if our conclusion is accepted. Assume that we have a PbTe system with $n$ molecules, and $\Psi$ is the eigenvector of valence band, $H\Psi = E_v\Psi$. The eigenvector can be written as $\Psi = \sum_i(-c_{is}\varphi_{is} + c_{ip}\varphi_{ip})$. At the *n*th site, we then substitute a Mg atom for a Pb atom. Considering a simple situation, Δε for the nth site is changed to –Δε, and the Hamiltonian is denoted as $H_d$. It is reasonable to construct an initial approximate wave-vector as $\Psi' = \sum_{i\neq n}(-c_{is}\varphi_{is} + c_{ip}\varphi_{ip}) + (c_{ns}\varphi_{ns} + c_{np}\varphi_{np})$. The eigenvector of $H_d$ can be obtained using iterative algorithm, $|\Psi_d\rangle = H_d^m|\Psi'\rangle$, where $m$ is iteration number. Doing some algebra calculations, we could have the following conclusions.

The angle of Ψ and Ψ' is $\cos\alpha = \langle\Psi|\Psi'\rangle = 1 - c_{ns}^2$. For $n \geq 3$, we obtain $\cos\alpha > \sqrt{3}/3$. This angle indicates that Ψ is the closest basic vector to Ψ' in the orthogonal space expanded by the eigenvectors of *H*, *i.e.*, Ψ' is a distorted wave-vector of the valence band Ψ. After the first iteration, the wave-function is $|H_d\Psi'\rangle$. We obtained $\langle\Psi'|H_d\Psi'\rangle$ less than $\langle\Psi|H_d\Psi'\rangle$. This means that the iteration wave-function is closer to Ψ', rather than Ψ, *i.e.*, Ψ' is a better initial wave-vector than Ψ. Meanwhile, the iteration energy $\sqrt{\langle H_d\Psi'|H_d\Psi'\rangle}$ is less than $E_v = \langle\Psi|H\Psi\rangle$, indicating the downward shift of valence band.

The above discussions show that the eigenvector $\Psi_d$ of the doped system tends to have a similar form to Ψ'. Ψ' indicates a bulk anti-bonding state combined with a defect bonding state. Meanwhile, the test wave function Ψ' decreases the eigenvalues of valence band. We then conclude that the BC effects of Mg dopant is due to the defect *s-p* bonding state, which offsets the bulk *s-p* anti-bonding state. The conclusion is illustrated in Fig. 3. According to the discussion of our model, the strength of a dopant *s-p* bonding state mainly depends on the distance of $\varepsilon_{D\text{-}s}$ and $\varepsilon_{Te\text{-}p}$. If the *s* orbital level of a dopant is as close as possible to Te-*p* orbital from the upward side, the dopant will cause a strong defect *s-p* bonding state, and lead to significant BC effects.

In Table 2, besides the experimental reported effective dopants of Mg, Mn, and Cd, we found that Be has an ideal s orbital level. We applied DFT calculations to examine this dopant. **Figure 6**(a-b) presents the calculated band structure for Be-doped SnTe, compared with that for Mg-doped SnTe. It may be seen that Be doping enlarges the band gap and decreases the $\Delta E_{L\text{-}\Sigma}$ significantly. We did not find any experimental reports on Be doping in PbTe or SnTe, maybe owing to its poisonousness. This example shows that our model causes band engineering in PbTe/SnTe to be designable.

Using our model, some unclarified experiments can be explained.

1) Enlarged band gap in Na-doped PbTe.[16] In PbTe, Na doping is often applied to increase the hole concentration.[3,9,38] Meanwhile, experimentalists have realized that Na doping also increases the band gap by about 0.05 eV,[16] even though this observation was not officially explained. Our model indicates that the enlarged band gap is due to the s orbital level of Na. Na-*s* is –4.8 eV, which is higher than Te-*p* of –9.54 eV, thus the introduced *s-p* bonding state can increase the band gap. On the other hand, the distance between Na-*s* and Te-*p* is large, and the BC effects of Na doping are not as significant as in Mg-doped PbTe.

2) Different efficiency of BC dopants. Although the dopants of Mg, Mn, Cd, and Hg all induced BC effects in PbTe/SnTe, experimentalists found that Cd and Hg are more efficient than Mg and Mn,[11,14,15,26,28,29,30,31,32,33] *i.e.*, similar BC effects requires lower Cd or Hg doping concentration than Mg or Mn doping. Table 2 shows that the *s* orbital level of Cd or Hg is lower than that of Mg or Mn, and therefore Cd or Hg doping has stronger BC effects. Note that the influence of crystal field is ignored in our model. We can simply tell that Cd is more efficient than Ca or Na, but it is difficult to tell Mg or Mn is more efficient.

3) BC effects induced by transition metals. Mn and Eu have the partly filled *d* or *f* shell, while they were found to be effective dopants for BC in PbTe/SnTe.[14,16,28,29,30,31] A problem is that whether the *d* or *f* electron influences the BC effects. Our model shows that the valence band is from the bulk *s-p* anti-bonding state. When a d or f orbital appears on Fermi level, it will introduce new defect levels, but not tune the bulk valence band dispersion, as reported in Cr- and Fe-doped PbTe.[64] Fortunately, Mn or Eu have the half occupied d or f shell, magnetic exchange interaction can split the d or f shell into the two sub-shells for spin-up and spin-down.

The two sub-shells are either fully occupied or fully empty, and thus the d or f orbital does not appear on Fermi level.[28,65] Therefore, the BC effects of Mn and Eu dopants are also from their *s* orbital. Researchers also observed BC effects in Yb-doped PbTe.[66] This problem is similar to that of Mn or Eu, since the whole *f* shell of Yb is occupied. Eu-*s* and Yb-*s* are assumed to be close to Lu-*s*, which is –5.42 eV in Harrison's Table.[62]

### 3.2. $\varepsilon_{Sn/Pb\text{-}s} < \varepsilon_{D\text{-}s} < \varepsilon_{Te\text{-}p}$: Resonant Level

As reported by Heremans *et al.*, Tl doping introduces RL in *p*-type PbTe, leading to the increased *ZT* from 0.7 to 1.5 at 773 K.[4] In 2013, Zhang *et al.* observed RL in In-doped SnTe, and the increase *ZT* is up to 1.1 at 873 K.[35]

According to our model, if the *s*-level of a dopant is between $\varepsilon_{Sn/Pb\text{-}s}$ and $\varepsilon_{Te\text{-}p}$, the dopant will introduce a stronger molecular anti-bonding state than that of in bulk. If this anti-bonding state is located at the edge of band gap and it is strong enough, a resonant level will appear. Compared with BC, RL is very elementselective. **Figure 7**(a) presents the partial DOS for Ga-, In-, Tl-doped SnTe. It is clear that only In doping causes a RL at the edge of band gap. Table 2 lists the *s* energy levels for Ga, In and Tl. These *s* energy levels are all higher than $\varepsilon_{Sn/Pb\text{-}s}$ but lower than $\varepsilon_{Te\text{-}p}$. Our theoretical explanation for the mechanism of RL is consistent with DFT results.[67]

Table 2 shows that the *s* energy levels of Ga, In and Tl has the order of $\varepsilon_{Ga\text{-}s} < \varepsilon_{In\text{-}s} < \varepsilon_{Tl\text{-}s}$. However, the partial DOS shows that the anti-bonding state by In doping is highest. This contradiction may be due to two reasons. The first reason is the lanthanide contraction effect. Owing to the poor shielding of nuclear charge by 4*f* electrons, some elements in the 6*th* period have smaller atomic radii than their counterparts in the 5*th* period. This effect may cause Tl-6*s* to be lower than In-5*s*. The second reason is crystal field. In the same group, crystal field usually affects the upper elements more than the lowers, thus $\varepsilon_{In\text{-}s}$ is upwards shifted more than $\varepsilon_{Tl\text{-}s}$.

RL requires a defect state to be close to band edge, and be strong enough to introduce a sharp peak. That is the reason why RL is very element-selective. For example, Ga doping in SnTe hardly improves the thermoelectric performance.[68] This feature is different from BC. It is more difficult to predict correct dopant for RL, even though our model can largely reduce the candidate list.

It is interesting that the RL dopants in PbTe and SnTe are different, and their band structure configurations are different. Fig. 7(b) presents the schematic band structures for band engineering in PbTe and SnTe. Tl doping in PbTe introduces a RL at the edge of band gap,[10] where the location of RL in In-doped SnTe is a little bit lower than band edge.[67] From the statements of Mahan *et al.*,[8] the higher background DOS is not favorite to RL. Thus the RL effects in Tl-doped PbTe are more significant than in Indoped SnTe. On the other hand, the band configuration of SnTe could bring other benefits. The mixture of the RL and heavy, light valence bands can cause the synergy of BC and RL.

If the Fermi level of a given system is located on the peak of the defect state, RL exhibits its strongest effects. When temperature increases, the RL effects will decrease for the upward shift of Fermi level. In SnTe, it was clearly shown that the RL effects at room temperature is more significant than those at higher temperature.[35] On the other hand, BC effects appear at high temperature for the non-zero distance between heavy and light bands. Thus the synergy of BC and RL may improve the thermoelectric performance within a wide temperature range.

According to our explanation, In doping is irreplaceable for the synergy of BC and RL in SnTe. Tan *et al.* have realized the synergy of BC and RL in In-Cd codoped SnTe and the maximum *ZT* of $Sn_{0.97}In_{0.015}Cd_{0.015}Te$ reaches 1.4 at 923 K.[69] Banik *et al.* obtained a high power factor of 31.4 $\mu Wcm^{-1}K^{-2}$ in $SnIn_{0.025}Ag_{0.025}Te_{1.05}$.[70] In these works RL and BC were equally treated, however, it might not be the most efficient way. BC is a tuning of the periodical band structure and it requires a relative high doping level, while RL is a distortion of DOS and it may largely decrease s owing to the reducing of carrier mobility at high doping level. To optimize the synergy, the two kinds of dopant concentration need to be carefully tuned. Recently, it was report that the synergy of BC and RL increases the Seebeck coefficient of $Sn_{0.905}In_{0.005}Mn_{0.11}Te$ from 20 to 116 $\mu VK^{-1}$ at room temperature.[71] By optimizing the position of Fermi level, the average *ZT* of $Sn_{0.958}In_{0.01}Cd_{0.03}Te$ was increased to 0.54 from 300 K to 850 K.[72]

### 3.3. $\varepsilon_{D\text{-}s} < \varepsilon_{Sn/Pb\text{-}s}$: Weak Band Convergence

If the *s*-level of a dopant is lower than $\varepsilon_{Sn/Pb\text{-}s}$, the dopant will also produce a molecular anti-bonding state, while its strength is weaker than that of in bulk. So the whole anti-bonding state of the doped system will be weaker than the undoped system, but such decreasing is very limited. In Table 2, we list the *s* energy levels of As, Sb, and Bi. Fig. 6(c) presents the calculated band structure of Sb doped SnTe. As may be seen, the valence band is slightly tuned. In principle, we may not observe significant BC effects in Sb doped system.

In PbTe, Sb and Bi doping cause a transition from *p*-type to *n*-type.[73] This is not surprising because Sb or Bi have one more valence electron than Sn. In the *n*-type PbTe, the valence band tuning hardly influence the thermoelectric performance. Thus we do not discuss the doping effects of Sb or Bi in PbTe. The transition from *p*-type to *n*-type was not observed in Sb- or Bi-doped SnTe. Meanwhile, the experiments on Sb- or Bi-doped SnTe are kind of controversial, for some authors did not observe band engineering effects,[68] but some observed.[74,75] These controversies will be discussed in next subsection.

### 3.4. $\varepsilon_{D\text{-}s} = \pm\infty$: Effects of Cation Vacancy

In the thermoelectric field, vacancy is often related phonon scattering. Some researchers introduced certain vacancies to reduce the lattice thermal conductivity,[36] whereas the influence of vacancy on band engineering was barely realized. In SnTe, it has been reported that vacancy plays an important role in the band engineering by Mg or Mn doping.[76]

Within our model, vacancy can be regarded as a dopant with *s* energy level of $\pm\infty$. A vacancy does not contribute a bonding or anti-bonding state. As shown in **Figure 8**(a), the different doping effects are schematically described. At $\varepsilon_{D\text{-}s} > \varepsilon_{Te\text{-}p}$, the bulk antibonding state is reduced by the offsetting of a defect bonding state as well as the removing of a Pb/Sn-Te molecular anti-bonding state, leading to strong BC effects. At $\varepsilon_{Sn/Pb\text{-}s} < \varepsilon_{D\text{-}s} < \varepsilon_{Te\text{-}p}$, the bulk anti-bonding state is enhanced by the difference between the stronger defect anti-bonding state and the weaker Sn/Pb-Te molecular anti-bonding state. At $\varepsilon_{D\text{-}s} < \varepsilon_{Sn/Pb\text{-}s}$, the bulk anti-bonding state is slightly reduced by the difference between the weaker defect anti-bonding state and the stronger Pb/Sn-Te molecular anti-bonding state. At $\varepsilon_{D\text{-}s} = \pm\infty$, the bulk antibonding state is reduced by the removing of a Pb/Sn-Te molecular anti-bonding state. So the strength of BC effects of vacancy should be between those of Sb and Mg.

Experimental measurements observed strong BC effects in $Sn_{1-x}Mn_xTe$ at $x \geq 0.03$.[28,29,30,31] However, DFT calculations failed to reproduce the observable BC effects at $x = 0.037$, while the expected BC effects appear at a higher doping concentration of $x = 0.074$.[30,31] This discrepancy was resolved by a later theoretical study. When a vacancy is introduced to the computational supercell, the expected BC effects was reproduced in $Sn_{1-x}Mn_xTe$ at $x = 0.037$.[76] Note that the hole concentration in SnTe is at a high level of $10^{20}$–$10^{21}$ cm$^{-3}$,[21,77] owing to the negative formation energy of $V_{Sn}^{2-}$.[78] According to our model, the mechanism of the BC effects by vacancy can be clearly seen from the illustration of Fig. 8(a). In Fig. 8(b-c), we present the calculated band structure for vacancies in PbTe and SnTe. As may be seen, the vacancy BC effects in PbTe are fairly obvious. In SnTe, the presented BC effects are not as significant as those for Mg or Be doping, but is stronger than those for Sb doping.

Another signal for the moderate BC effects of vacancy is from the studies on Sb-doped SnTe.[68,75] Sb doping in SnTe has different behaviors from in PbTe. Sb doping in PbTe induced an *n*-type semiconductor, while $Sn_{1-x}Sb_xTe$ is still *p*-type semiconductor until $x = 0.3$.[73] At high doping concentration, Sb doping improved Seebeck coefficients and power factors of SnTe.[75] Compared with the results of Sn self-compensate experiment, it is clear that the effects of Sb doping are not from hole concentration tuning.[26,32,79,80,81,82,83] As discussed in the previous subsection, Sb doping has a weak BC effects, which should not be so significant as experimental observed. Considering the *p*-type behavior, it is reasonable to assume that Sb doping induced some Sn vacancies. As shown in Fig. 8(c), vacancies lead to stronger BC effects than Sb. The combined BC effects of vacancy and Sb may explain the improvement of Seebeck coefficients and power factors.

## 4. Conclusion

The phase diagram for the band engineering in *p*-type PbTe/SnTe is firstly established, starting from the fundamental understanding of band structure. It is disclosed that the valence band dispersion is mainly determined by the anti-bonding of cation-*s* and anion-*p* orbitals. The constructed *s*-*p* bonding model indicates that the doping effects mainly depends on an inherent parameter of doping element: the s orbital energy level, $\varepsilon_{D\text{-}s}$. At $\varepsilon_{D\text{-}s} > \varepsilon_{Te\text{-}p}$, the dopant contributes a bonding state to the valence band, and it partly offsets the anti-bonding state from the bulk, leading to the enlarged band gap and the decreased distance between light and heavy bands. At $\varepsilon_{Sn/Pb\text{-}s} < \varepsilon_{D\text{-}s} < \varepsilon_{Te\text{-}p}$, the dopant contributes an enhanced anti-bonding state, which may lead to the RL effects depending on the position of defect state. At $\varepsilon_{D\text{-}s} < \varepsilon_{Sn/Pb\text{-}s}$, the dopant contributes a reduced anti-bonding state, leading to slight BC effects. Interestingly, the *s*-*p* model indicates that a Pb/Sn vacancy can be regarded as a $\varepsilon_{D\text{-}s} = \pm\infty$ dopant, which induces the moderate BC effects. The established phase diagram presents a self-consistent explanation to the band engineering in *p*-type PbTe/SnTe, and it can predict new effective dopants. As demonstrated by band structure calculations, the predicted element, Be, is a very efficient dopant for band engineering.

As for a covalent bonding semiconductor, its valence band or conduction band is often from the anti-bonding or bonding state between the nearest neighbors. So our analytical method is not only suitable to PbTe/SnTe. The key step of the analysis is to understand the *elementary* band structure, *i.e.*, how the band dispersion depends on atomic orbitals. In this work, we mainly consider the influence of atomic energy levels, because they give the diagonal elements of Hamiltonian. For a lightly doped system, this approximation is always good enough. In a heavily doped system, the crystal group and lattice constant might be significantly changed, such as in $Pb_{1-x}Se_xTe$[3] and $Mg_2Si_{1-x}Sn_x$.[84] In those situations, the effects of crystal field and hopping integrals need to carefully considered, for they are sensitive to crystal structures. In summary, this work present an analytical method to understand and design band engineering in PbTe/SnTe, and this method can be adopted to more thermoelectric materials.

**Acknowledgements**

This work was supported by the National Natural Science Foundation of China (11404350, and 11234012), Natural Science Foundation of Zhejiang Province (LY18A040008, and LY18E020017) and Zhejiang Provincial Science Foundation for Distinguished Young Scholars (LR16E020001).

Received: ((will be filled in by the editorial staff))
Revised: ((will be filled in by the editorial staff))
Published online: ((will be filled in by the editorial staff))

**References**

[1] L. Bell, *Science* **2008**, 321, 1457.

[2] G. A. Slack, in *CRC Handbook of Thermoelectrics*, (edited by D. M. Rowe), CRC press, Boca Raton, FL **1995**, p. 407.

[3] Y. Pei, X. Shi, A. LaLonde, H. Wang, L. Chen, G. J. Snyder, *Nature* **2011**, 473, 66.

[4] J. P. Heremans, V. Jovovic, E. S. Toberer, A. Saramat, K. Kurosaki, A. Charoenphakdee, S. Yamanaka, G. J. Snyder, *Science* **2008**, 321, 554.

[5] Y. Pei, A. D. LaLonde, H. Wang, G. J. Snyder, *Energy Environ. Sci.* **2012**, 5, 7963.

[6] G. Tan, L.-D. Zhao, M. G. Kanatzidis, *Chem. Rev.* **2016**, 116, 12123.

[7] W. G. Zeier, A. Zevalkink, Z. M. Gibbs, G. Hautier, M. G. Kanatzidis, G. J. Snyder, *Angew. Chem. Int. Ed.* **2016**, 55, 6826.

[8] G. D. Mahan and J. O. Sofo, *Proc. Natl. Acad. Sci. U. S. A.* **1996**, 93, 7436.

[9] Y. Pei, H. Wang, G. J. Snyder, *Adv. Mater.* **2012**, 24, 6125.

[10] J. P. Heremans, B. Wiendlocha, A. M. Chamoire, *Energy Environ. Sci.* **2012**, 5, 5510.

[11] Y. Pei, A. D. LaLonde, N. A. Heinz, X. Shi, S. Iwanaga, H. Wang, L. Chen, G. J. Snyder, *Adv. Mater.* **2011**, 23, 5674.

[12] K. Biswas, J. He, G. Wang, S.-H. Lo, C. Uher, V. P. Dravid, M. G. Kanatzidis, *Energy Environ. Sci.* **2011**, 4, 4675.

[13] G. Tan, F. Shi, S. Hao, L.-D. Zhao, H. Chi, X. Zhang, C. Uher, C. Wolverton, V. P. Dravid, M. G. Kanatzidis, *Nat. Commun.* **2016**, 7, 12167.

[14] Y. Pei, H. Wang, Z. M. Gibbs, A. D. LaLonde, G. J. Snyder, *NPG Asia Mater.* **2012**, e4, 428.

[15] Y. Pei, A. D. LaLonde, N. A. Heinz, G. J. Snyder, *Adv. Energy Mater.* **2012**, 2, 670.

[16] Z. Chen, Z. Jian, W. Li, Y. Chang, B. Ge, R. Hanus, J. Yang, Y. Chen, M. Huang, G. J. Snyder, Y. Pei, *Adv. Mater.* **2017**, 29, 1606768.

[17] J. D. König, M. D. Nielsen, Y.-B. Gao, M. Winkler, A. Jacquot, H. Böttner, J. P. Heremans, *Phys. Rev. B* **2011**, 84, 205126.

[18] E. P. Skipetrov, L. A. Skipetrova, A. V. Knotko, E. I. Slynko, V. E. Slynko, *J. Appl. Phys.* **2014**, 115, 133702.

[19] M. D. Nielsen, E. M. Levin, C. M. Jaworski, K. Schmidt-Rohr, J. P. Heremans, *Phys. Rev. B* **2012**, 85, 045210.

[20] K. Biswas, J. He, I. D. Blum, C.-I. Wu, T. P. Hogan, D. N. Seidman, V. P. Dravid, M. G. Kanatzidis, *Nature* **2013**,489, 414.

[21] R. F. Brebrick, A. J. Strauss, *Phys. Rev.* **1963**, 131, 104.

[22] R. Brebrick, *J. Phys. Chem. Solids* **1963**, 24, 27.

[23] L. M. Rogers, *J. Phys. D: Appl. Phys.* **1968**, 1, 845.

[24] W. Li, Y. Wu, S. Lin, Z. Chen, J. Li, X. Zhang, L. Zheng, Y. Pei, *ACS Energy Lett.* **2017**, 2, 2349.

[25] R. Moshwan, L. Yang, J. Zou, Z.-G. Chen, *Adv. Funct. Mater.* **2017**, 1703278.

[26] A. Banik, U. S. Shenoy, S. Anand, U. V. Waghmare, K. Biswas, *Chem. Mater.* **2015**, 27, 581.

[27] R. Al Rahal Al Orabi, N. A. Mecholsky, J. Hwang, W. Kim, J.-S. Rhyee, D. Wee, M. Fornari, *Chem. Mater.* **2016**, 28, 376.

[28] J. He, X. Tan, J. Xu, G.-Q. Liu, H. Shao, Y. Fu, X. Wang, Z. Liu, J. Xu, H. Jiang, J. Jiang, *J. Mater. Chem. A* **2015**, 3, 19974.

[29] G. Tan, F. Shi, S. Hao, H. Chi, T. P. Bailey, L.-D. Zhao, C. Uher, C. Wolverton, V. P. Dravid, M. G. Kanatzidis, *J. Am. Chem. Soc.* **2015**, 137, 11507.

[30] H. Wu, C. Chang, D. Feng, Y. Xiao, X. Zhang, Y. Pei, L. Zheng, D. Wu, S. Gong, Y. Chen, J. He, M. G. Kanatzidis, L.-D. Zhao, *Energy Environ. Sci.* **2015**, 8, 3298.

[31] W. Li, Z. Chen, S. Lin, Y. Chang, B. Ge, Y. Chen, Y. Pei, *J. Materiomics* **2015**, 1, 307.

[32] G. Tan, L.-D. Zhao, F. Shi, J. W. Doak, S.-H. Lo, H. Sun, C. Wolverton, V. P. Dravid, C. Uher, M. G. Kanatzidis, *J. Am. Chem. Soc.* **2014**, 136, 7006.

[33] J. He, J. Xu, G.-Q. Liu, H. Shao, X. Tan, Z. Liu, J. Xu, H. Jiang, J. Jiang, *RSC Adv.* **2016**, 6, 32189.

[34] G. Tan, F. Shi, J. W. Doak, H. Sun, L.-D. Zhao, P. Wang, C. Uher, C. Wolverton, V. P. Dravid, M. G. Kanatzidis, *Energy Environ. Sci.* **2015**, 8, 267.

[35] Q. Zhang, B. Liao, Y. Lan, K. Lukas, W. Liu, K. Esfarjani, C. Opeil, D. Broido, G. Chen, Z. Ren, *Proc. Natl. Acad. Sci. U. S. A.* **2013**, 110, 13261.

[36] G. Tan, W. G. Zeier, F. Shi, P. Wang, G. J. Snyder, V. P. Dravid, M. G. Kanatzidis, *Chem. Mater.* **2015**, 27, 7801.

[37] W. Li, L. Zheng, B. Ge, S. Lin, X. Zhang, Z. Chen, Y. Chan, Y. Pei, *Adv. Mater.* **2017**, 29, 1605887.

[38] Y. Pei, N. A. Heinz, A. LaLonde, G. J. Snyder, *Energy Environ. Sci.* **2011**, **4**, 3640.

[39] Y. Gelbstein, Z. Dashevsky, Y. George, M. P. Dariel, *Int. Conf. Thermoelectr. 25th*, **2006**, 418.

[40] J. Androulakis, C.-H. Lin, H.-J. Kong, C. Uher, C.-I. Wu, T. Hogan, B. A. Cook, T. Caillat, K. M. Paraskevopoulos, M. G. Kanatzidis, *J. Am. Chem. Soc.* **2007**, 129, 9780.

[41] G. Braunstein, G. Dresselhaus, J. Heremans, D. Partin, *Phys. Rev. B* **1987**, 35, 1969.
[42] D. L. Partin, C. M. Thrush, B. M. Clemens, *J. Vac. Sci. Technol. B* **1987**, 5, 686.
[43] F. F. Aliev, H. A. Hasanov, *Inorg. Mater.* **2011**, 47, 853.
[44] B. A. Volkov, L. I. Ryabova, D. R. Khokhlov, *Phys.-Usp.* **2002**, 45, 819.
[45] S. Takaoka, T. Itoga, K. Murase, *Jpn. J. Appl. Phys.* **1984**, 23, 216.
[46] A. D. LaLonde, Y. Pei, G. J. Snyder, *Energy Environ. Sci.* **2011**, 4, 2090.
[47] A. Bruno, J. P. Lascaray, M. Averous, G. Fillion, J. F. Dumas, *Phys. Rev. B* **1988**, 37, 1186.
[48] D. L. Partin, *J. Appl. Phys.* **1985**, 57, 1997.
[49] J. B. Conklin, L. E. Johnson, G. W. Pratt, *Phys. Rev.* **1965**, 137, A1282.
[50] J. O. Dimmock, G. B. Wright, *Phys. Rev.* **1964**, 135, A821.
[51] W. Paui, R. V. Jones, *Proc. Phys. Soc. Sect. B* **1953**, 66, 194.
[52] I. I. Zasavitskii, E. A. de Andrada e Silva, E. Abramof, P. J. McCann, *Phys. Rev. B* **2004**, 70, 115302.
[53] S. V. Ovsyannikov, V. V. Shchennikov, A. Y. Manakov, A. Y. Likhacheva, Y. S. Ponosov, V. E. Mogilenskikh, A. P. Vokhmyanin, A. I. Ancharov, E. P. Skipetrov, *Phys. Status Solidi B* **2009**, 246, 615.
[54] W. H. Strehlow, E. L. Cook, *J. Phys. Chem. Ref. Data* **1973**, 2, 163.
[55] A. Svane, N. E. Christensen, M. Cardona, A. N. Chantis, M. van Schilfgaarde, T. Kotani, *Phys. Rev. B* **2010**, 81, 245120.
[56] A. Alkauskas, P. Broqvist, A. Pasquarello, *Phys. Status Solidi B* **2011**, 248, 775.
[57] T. Shimazaki, Y. Asai, *Chem. Phys. Lett.* **2008**, 466, 91.
[58] T. Shimazaki, T. Nakajima, *J. Chem. Phys.* **2015**, 142, 074109.
[59] J. H. Skone, M. Govoni, G. Galli, *Phys. Rev. B* **2014**, 89, 195112.
[60] L. Hedin and S. Lundqvist, *Solid State Phys.* **1970**, 23, 1.
[61] M. S. Hybertsen, S. G. Louie, *Phys. Rev. B* **1986**, 34, 5390.
[62] W. A. Harrison, *Elementary Electronic Structure*, World Scientific, Singapore, **2003**.
[63] S.-H. Wei, A. Zunger, *Phys. Rev. B* **1997**, 55, 13605.
[64] Y. I. Ravich, B. A. Efimova, I. A. Smirnov, *Semiconducting Lead Chalcogenides*, Plenum Press, New York, **1970**.
[65] X. Tan, H. Shao, T. Hu, G.-Q. Liu, S.-F. Ren, *J. Phys.: Condens. Matter* **2015**, 27, 095501.
[66] Z. Jian, Z. Chen, W. Li, J. Yang, W. Zhang, Y. Pei, *J. Mater. Chem. C* **2015**, 3, 12410.
[67] X. J. Tan, G. Q. Liu, J. T. Xu, H. Z. Shao, J. Jiang, H. C. Jiang, *Phys. Chem. Chem. Phys.* **2016**, 18, 20635.

[68] L.-D. Zhao, X. Zhang, H. Wu, G. Tan, Y. Pei, Y. Xiao, C. Chang, D. Wu, H. Chi, L. Zheng, S. Gong, C. Uher, J. He, M. G. Kanatzidis, *J. Am. Chem. Soc.* **2016**, 138, 2366.

[69] G. Tan, F. Shi, S. Hao, H. Chi, L.-D. Zhao, C. Uher, C. Wolverton, V. P. Dravid, M. G. Kanatzidis, *J. Am. Chem. Soc.* **2015**, 137, 5100.

[70] A. Banik, U. S. Shenoy, S. Saha, U. V. Waghmare, K. Biswas, *J. Am. Chem. Soc.* **2016**, 138, 13068.

[71] L. Wang, X. Tan, G. Liu, J. Xu, H. Shao, B. Yu, H. Jiang, S. Yue, J. Jiang, *ACS Energy Lett.* **2017**, 2, 1203.

[72] X. Tan, X. Tan, G. Liu, J. Xu, H. Shao, H. Hu, M. Jin, H. Jiang, J. Jiang, *J. Mater. Chem. C* **2017**, 5, 7504.

[73] L. D. Borisova, *Phys. Status Solidi A* **1979**, 53, K19.

[74] Z. Zhou, J. Yang, Q. Jiang, Y. Luo, D. Zhang, Y. Ren, X. He, J. Xin, *J. Mater. Chem. A* **2016**, 4, 13171.

[75] A. Banik, B. Vishal, S. Perumal, R. Datta, K. Biswas, *Energy Environ. Sci.* **2016**, 9, 2011.

[76] X. J. Tan, H. Z. Shao, J. He, G. Q. Liu, J. T. Xu, J. Jiang, H. C. Jiang, *Phys. Chem. Chem. Phys.* **2016**, 18, 7141.

[77] V. P. Vedeneev, S. P. Krivoruchko, E. P. Sabo, *Semiconductors* **1998**, 32, 241.

[78] N. Wang, D. West, J. Liu, J. Li, Q. Yan, B.-L. Gu, S. B. Zhang, W. Duan, *Phys. Rev. B* **2014**, 89, 045142.

[79] J. He, J. Xu, G.-Q. Liu, H. Shao, X. Tan, Z. Liu, J. Xu, H. Jiang, J. Jiang, *RSC Adv.* **2016**, 6, 32189.

[80] M. Zhou, Z. M. Gibbs, H. Wang, Y. Han, C. Xin, L. Li, G. J. Snyder, *Phys. Chem. Chem. Phys.* **2014**, 16, 20741.

[81] M. Zhou, Z. M. Gibbs, H. Wang, Y. Han, L. Li, G. J. Snyder, *Appl. Phys. Lett.* **2016**, 109, 042102.

[82] L. Zhang, J. Wang, Z. Cheng, Q. Sun, Z. Li, S. Dou, *J. Mater. Chem. A* **2016**, 4, 7936.

[83] X. Wang, K. Guo, I. Veremchuk, U. Burkhardt, X. Feng, J. Grin, J. Zhao, *J. Rare Earth.* **2015**, 33, 1175.

[84] W. Liu, X. Tan, K. Yin, H. Liu, X. Tang, J. Shi, Q. Zhang, C. Uher, *Phys. Rev. Lett.* **2012**, 108, 166601.

**Figure 1.** Schematic representation of band engineering in *p*-type thermoelectric materials: (a) convergence of two valence bands, and (b) distortion in density of state.

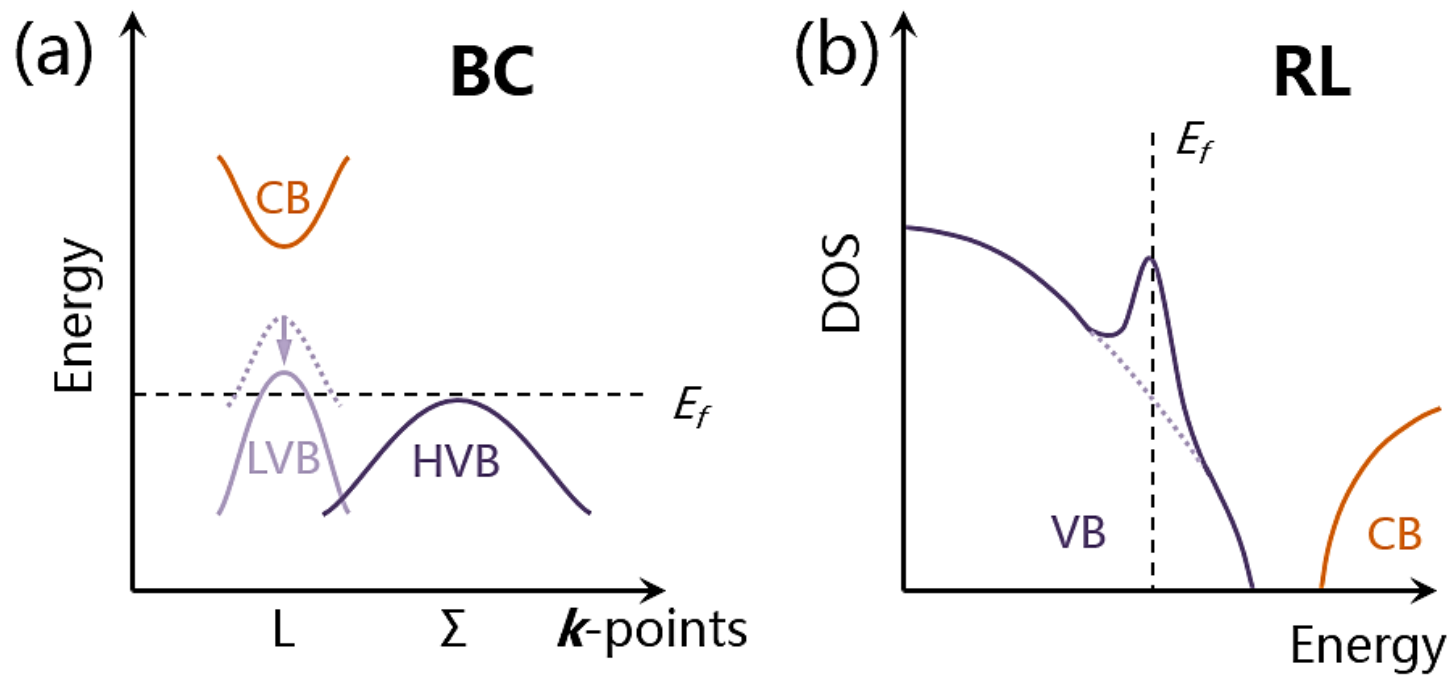

**Figure 2.** Calculated density of state and band structure for PbTe: (a) partial density of state, (b-d) band characters for Te-*p*, Pb-*s*, and Pb-*p* orbitals, respectively.

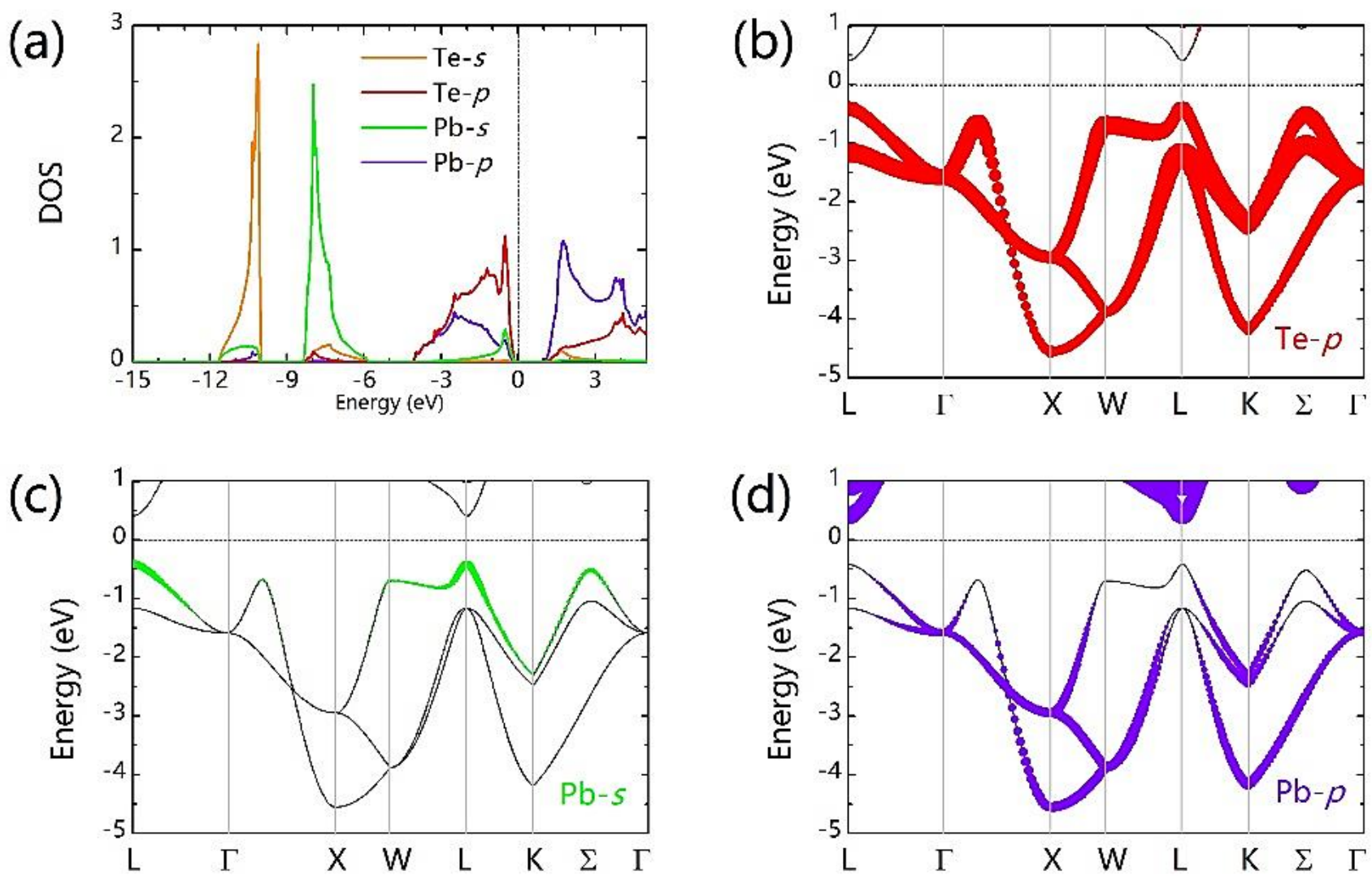

**Figure 3.** Calculated band feature by the *s*-*p* model: (a) reconstructed valence band dispersion, (b) the energy difference between $E_L$ and $E_\Sigma$ as a function of Δε and $t_{sp\sigma}$ for PbTe, and (c) reconstructed band dispersion for MgTe. The Pb-*s* and Mg-*s* character in DFT band structure is denoted by the blue dots.

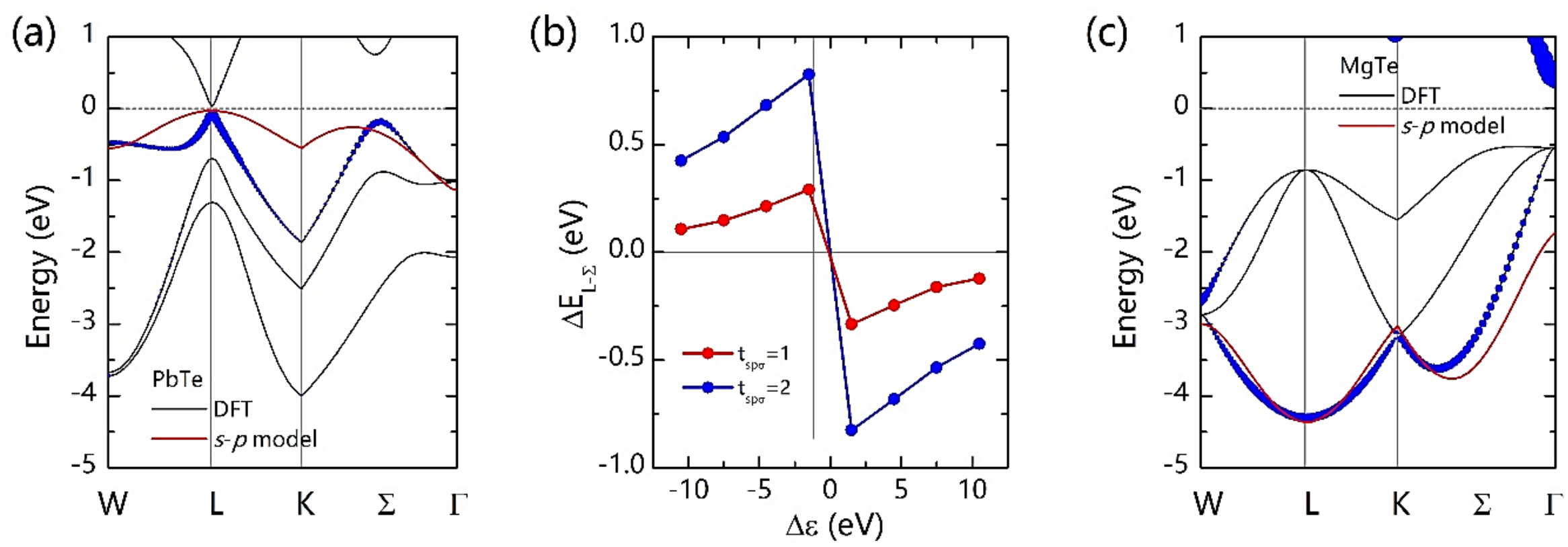

**Figure 4.** Established phase diagram for the band engineering in *p*-type PbTe/SnTe.

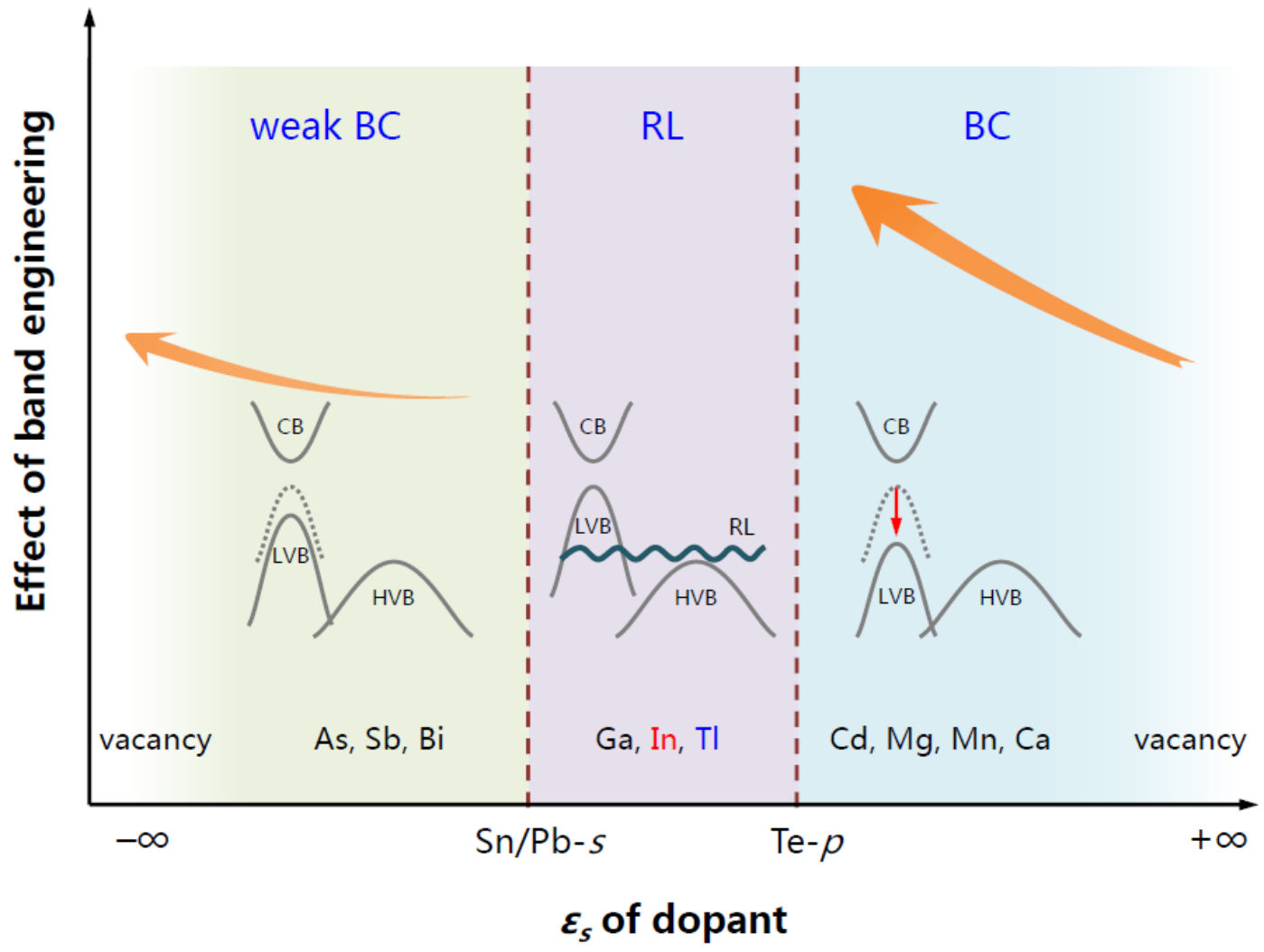

**Figure 5.** Band convergence increased Seebeck coefficient as a function of hole concentration *n* in *p*-type PbTe (left) and SnTe (right) at 300 K.

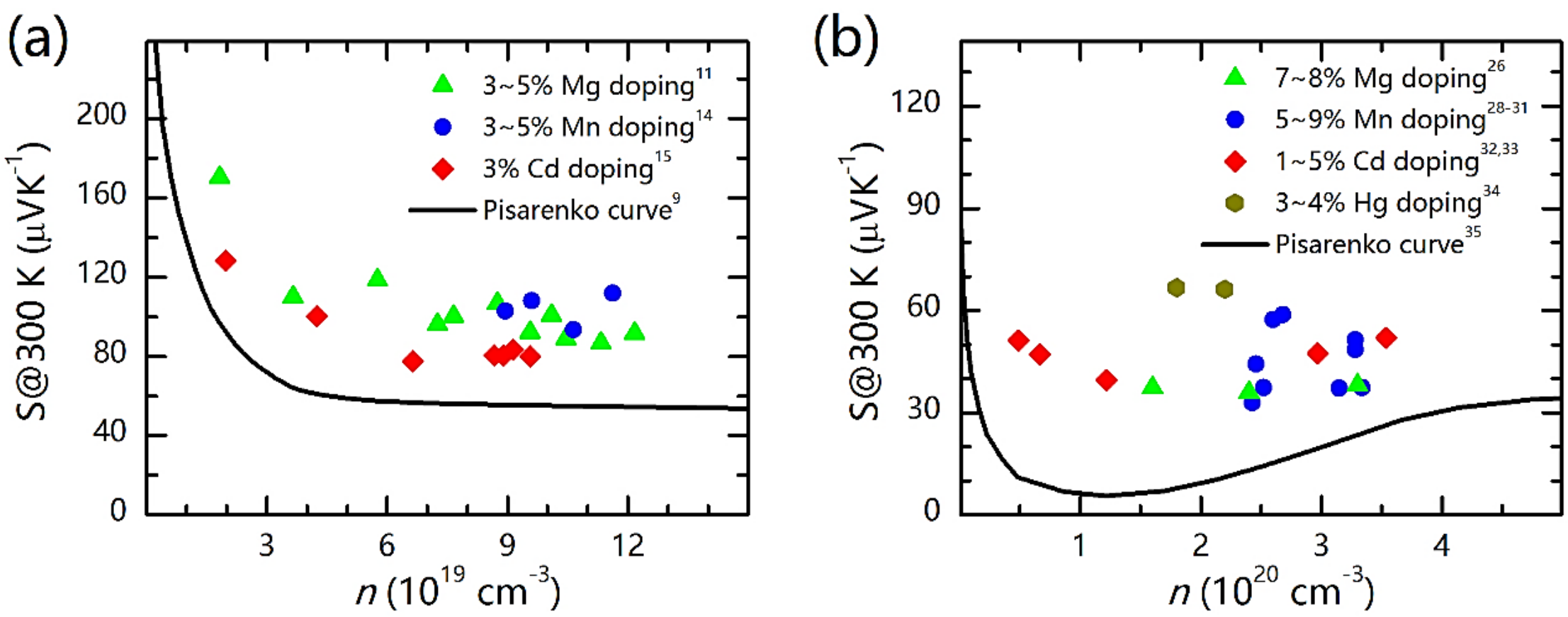

**Figure 6.** Calculated band structure for SnTe with 7.4% Be, Mg, or Sb doping.

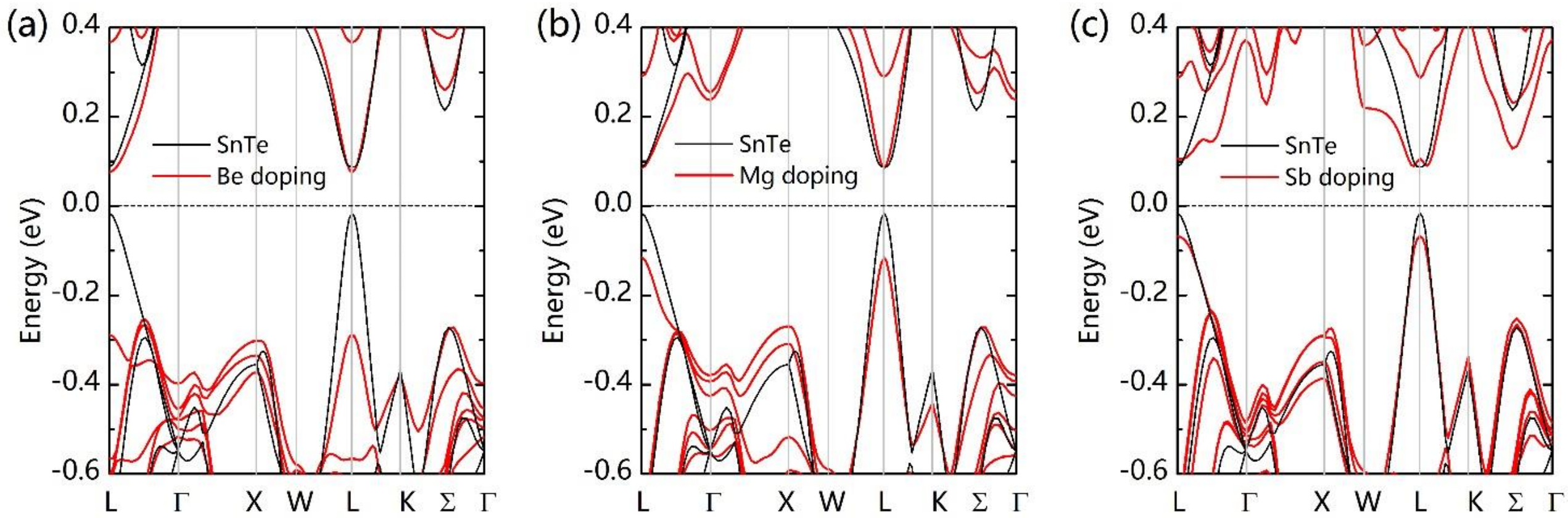

**Figure 7.** (a) Calculated partial density of state for SnTe with 3.7% Ga, In, or Tl doping, and schematic band engineering of (b) PbTe and (c) SnTe for describing the synergy of band convergence of the resonant level.

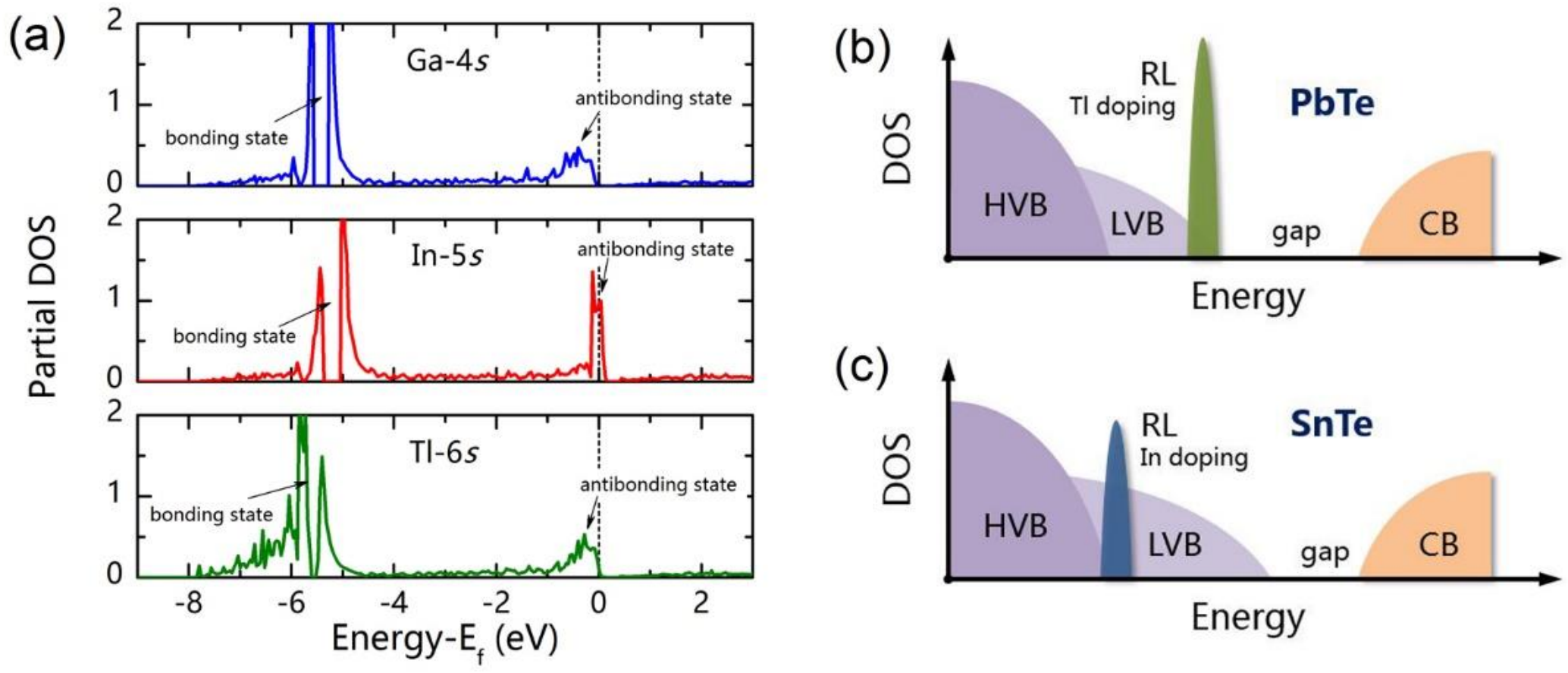

**Figure 8.** (a) Schematic representation of different band engineering effects for PbTe (SnTe), and calculated band structure for (b) PbTe and (c) SnTe with 7.4% cation vacancy.

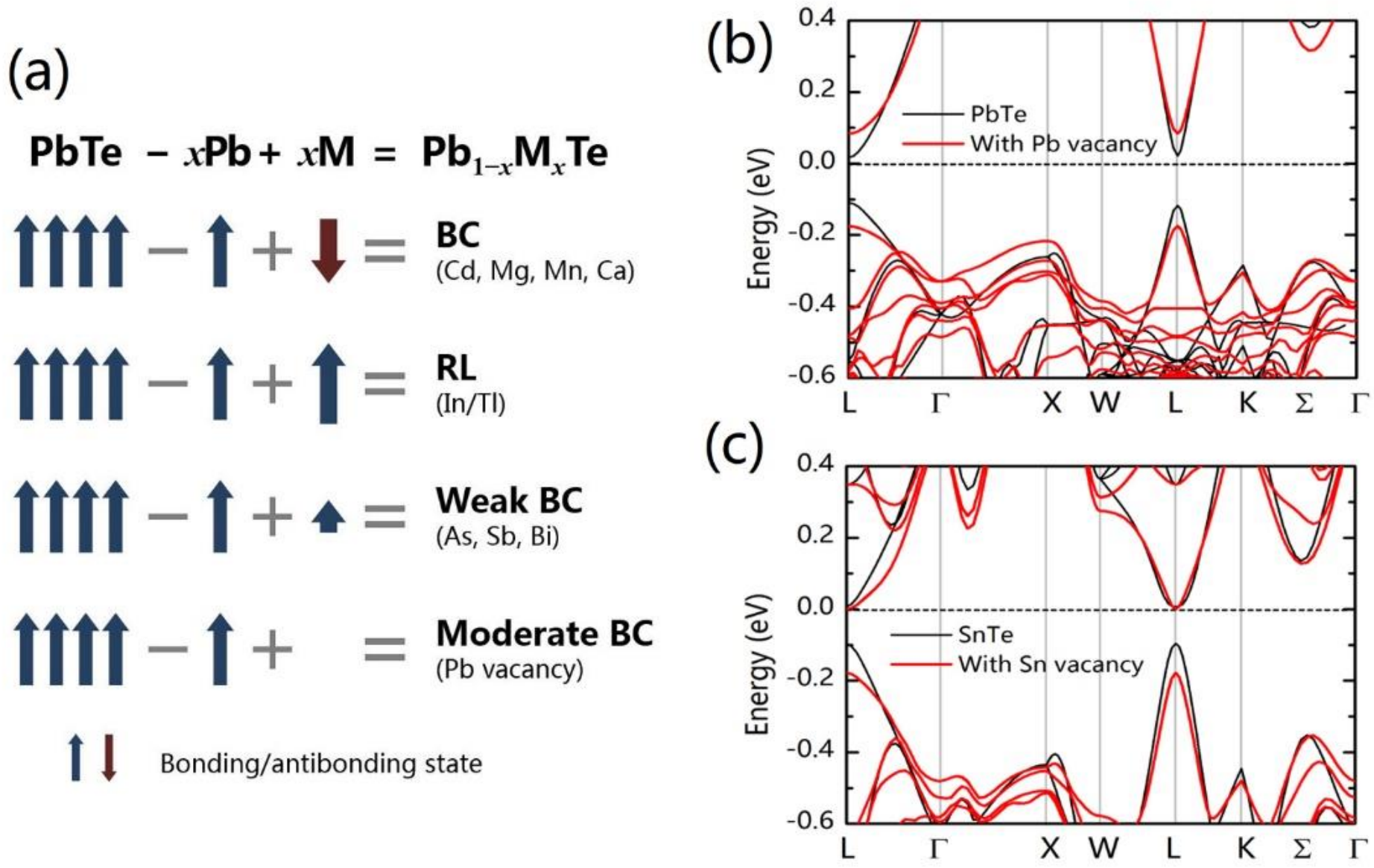

**Table 1.** Atomic energy levels for the orbitals of tin and lead chalcogenides, and reported band gap $E_g$ and energy separation between two valence bands $\Delta E_{L\text{-}\Sigma}$ for lead chalcogenides.

| | $\varepsilon_s$ (eV)[62] | $\varepsilon_p$ (eV)[62] | | $E_g$ (eV) exp.[54] | $\Delta E_{L\text{-}\Sigma}$ (eV) GW[55] |
|---|---|---|---|---|---|
| S | –24.0 | –11.6 | PbTe | 0.19 | 0.21 |
| Se | –22.9 | –10.7 | PbSe | 0.165 | 0.52 |
| Te | –19.1 | –9.5 | PbS | 0.286 | 0.66 |
| Sn | –13.0 | –6.8 | | | |
| Pb | –12.5 | –6.5 | | | |

**Table 2.** *s* orbital energy levels for the element with weak band convergence effects. The data are from reference [62].

| BC | | RL | | Weak BC | |
|---|---|---|---|---|---|
| Element | $\varepsilon_s$ (eV) | Element | $\varepsilon_s$ (eV) | Element | $\varepsilon_s$ (eV) |
| Na | –4.96 | *Te($\varepsilon_p$)* | *–9.54* | *Pb* | *–12.5* |
| Ca | –5.32 | Tl | –9.83 | *Sn* | *–13.0* |
| Lu | –5.42 | In | –10.14 | Bi | –15.19 |
| Mn | –6.84 | Ga | –11.55 | Sb | –16.03 |
| Mg | –6.89 | | | As | –18.92 |
| Hg | –7.10 | | | | |
| Cd | –7.21 | | | | |
| Zn | –7.96 | | | | |
| Be | –8.42 | | | | |

**Title: Establishing phase diagram for the band engineering in *p*-type PbTe/SnTe from elementary electronic structure understanding**

*Xiaojian Tan, Guoqiang Liu,* Jingtao Xu, Hezhu Shao, Haochuan Jiang, and Jun Jiang**

**Starting from the elementary understanding of the band structure**, the established phase diagram presents a self-consistent explanation to the band engineering in *p*-type PbTe/SnTe, and it can predict new effective dopants from the periodic table directly. This work will open a new thought to understand and design band engineering in more thermoelectric materials.

ToC figure (55 mm broad × 50 mm high)

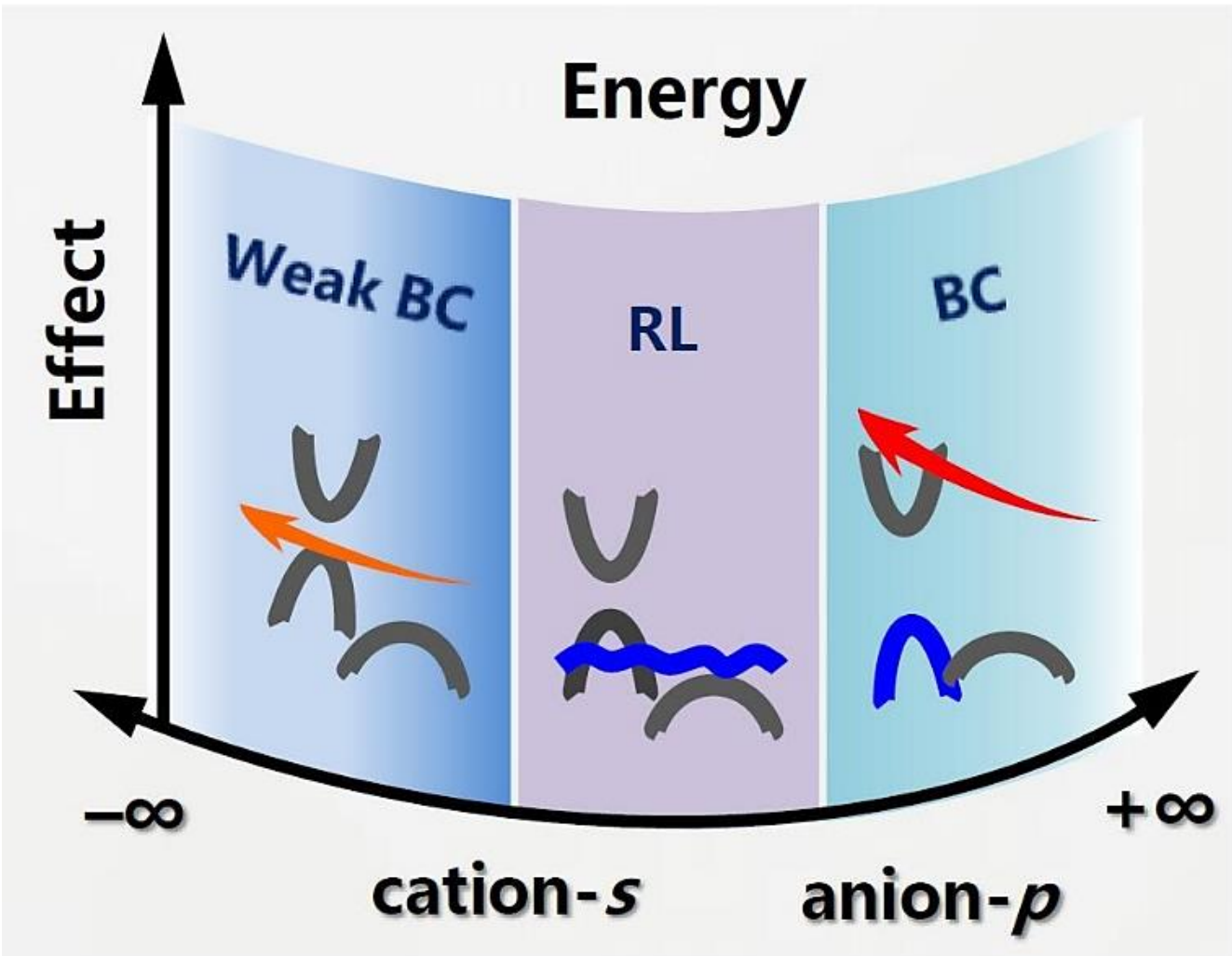